 \newlength\smallfigwidth
 \documentclass[aps,prb,twocolumn,floatfix,showpacs,amsmath,amssymb,superscriptaddress]{revtex4-1}

\def\ba{\begin{eqnarray}}
\def\ea{\end{eqnarray}}
\def\be{\begin{equation}}
\def\ee{\end{equation}}

\usepackage{graphicx}
\usepackage[hypertex]{hyperref}
 \usepackage[usenames,dvipsnames]{pstricks}
 \usepackage{epsfig}
 \usepackage{pst-grad} % For gradients
 \usepackage{pst-plot} % For axes
\begin{document}

%\preprint{UFV}

\title{Geometrically induced electric polarization in conical topological insulators}

\author{J.\ M.\ Fonseca}
\email{jakson.fonseca@ufv.br}
\affiliation{Grupo de F\'{\i}sica/ICET/CRP, Universidade Federal de Vi\c cosa,
Rodovia MG-230 Km 7, Cep 38810-000, Rio Parana\'iba, Minas Gerais, Brazil.}
\affiliation{Departamento de F\'isica, Universidade Federal de Vi\c cosa,
Cep 36570-000, Vi\c cosa, Minas Gerais, Brazil}
\author{W.\ A.\ Moura-Melo}
\email{winder@ufv.br}
\author{A.\ R.\ Pereira}
\email{apereira@ufv.br}
\affiliation{Departamento de F\'isica, Universidade Federal de Vi\c cosa,
Cep 36570-000, Vi\c cosa, Minas Gerais, Brazil}

\date{\today}

\begin{abstract}

We study the topological magnetoelectric effect on a conical topological insulator when a point charge $q$ 
is near the cone apex. The Hall current induced on the cone surface and the image 
charge configuration are determined. We also study a kind of gravitational Aharonov-Bohm effect in this geometry and realize a phase diference betwen the components of the wavefunctions (spinors) upon closed parallel transport around the (singular) 
cone tip. Concretely, a net current flowing towards cone apex (or botton) shows up, yielding electric polarization of the conical topological insulator. Such an effect may be detected, for instance, by means of the net accumulated Hall charge near the apex. Once it depends only on the geometry of the material (essetially, the cone apperture angle) this may be faced as a microscopic scale realization of (2+1)-dimensional Einstein gravity.
\end{abstract}
%
%73.43.-f quantum hall effects
%78.20.Ls: Magneto-optical effects
%03.65.Vf: Phases: geometric; dynamic or topological
%73.20.-r: Electron states at surfaces and interfaces
%75.30.Hx: Magnetic impurity interactions
%85.75.-d: Magnetoelectronics; spintronics:
%04.20.-q: Classical general relativity
%

\maketitle

\indent  

Topological insulators (TI's) are a new class of quantum matter with an insulating bulk gap
generated by strong spin-orbit coupling and gapless matallic edge (2D TI's) 
or surface (3D TI's) states characterized by a topological index associated 
with the Bloch wave function in the momentum space. At present, this behaviour has been observed
in $Bi_xSb_{1-x}$  alloys and in $Bi_2Se_3$, $Bi_2Te_3$ and $Sb_2Te_3$ crystals \cite{Hsieh 2008, Chen, Xia} 
and it has been predicted to occur in ternary intermetallic Heusler compounds \cite{Chadov, Wray}.
The physical manifestation of the topological order comes in the form of protected, gapless, 
surface states that are robust against damaging the surface by chemical or mechanical means,
altering its shape or orientation  with respect to the crystal lattice, or even disordering the bulk, 
as long as such changes are applied in moderation\cite{TI review}.

These properties are verified in those systems that respect time reversal symmetry (TRS), say, nonmagnetic TI's at zero applied magnetic field. Actually, whenever TRS is broken, even by a weak 
magnetic perturbation, the spectrum of the topologically protected surface states acquire mass gap.
As a result, the system becomes fully insulating,
both at the bulk and on the surface. However, the  surface is not an ordinary 
insulator; it is rather a quantum Hall insulator with properties similar to those of the familiar 
quantum Hall systems realized in two-dimensional electron gas \cite{y. l. chen}.
In this case, the topological properties of the fully gapped insulator are characterized
by a novel topological magnetoelectric effect. A tunable energy gap at the surface Dirac point 
provides a means to control the surface electric transport, what is important for actual applications.
The experimental observation of massive Dirac Fermion on the 
surface of a magnetically doped TI was reported in Ref.\cite{y. l. chen} where the authors introduced 
magnetic dopants into 3D TI $Bi_2Se_3$ and observed the massive Dirac fermion state by 
angle-resolved photoemission.

By virtue of all that, the investigation of how topological insulator phase behave whenever coexisting with extra orders, such as magnetism, superconductivity etc, comes to be a necessity for real TI's applications. In this context, the authors of Ref.\cite{nomura and nagaosa 2010} have recently considered a thin ferromagnetic film, with spin textures such as domain walls and vortices, deposited on TI surface. The interaction with the TI electronic states has render these magnetic textures electrically charged. Other magnetic-TI interacting systems comprise magnetization dynamics induced by TI surface\cite{Garate,vozmediano}, magnetic impurities effects on Dirac electronic states \cite{Liu}. Thus, TI physical properties are dramatic if an energy gap is induced. For instance, when this is provided by a magnetic
field (or due to the proximity to a magnetic material) the resulting system behaves like quantum Hall state along with a topological magnetoelectric effect, in such a way that an electric (magnetic) field generates a topological contribution to the magnetization (polarization), with a universal constant of proportionality quantized in odd multiples of the
fine-structure constant $\alpha=e^2/\hbar c$ \cite{Qi, Essim}. On the other hand,  whenever coupled to a superconductor leads to novel states supporting Majorana fermions, which may provide a new venue for realizing topological quantum computation \cite{Fu}.

The TI effective model is obtained by projecting the bulk Hamiltonian onto
the surface states. In the simplest case of a single Dirac point, to the leading order 
in $\vec k$ (low energy spectrum), the surface states are well described by the Dirac-like Hamiltonian for a massless particle 
\cite{LiuHamiltonians}
\begin{equation}\label{Dirac Hamiltonian}
H_{\rm surf}=-i\hbar v_F \vec \sigma \cdot \vec \nabla\,,
\end{equation}
where $v_{F}$ is the Fermi velocity, while $\vec{\sigma}=(\sigma_{x}, \sigma_{y})$ are the $2D$ Pauli matrices.

Therefore, the surface states of a $3D$ TI are described quite similarly to their counterparts lying on a single graphene sheet \cite{Castro-Neto}. However, due to spin and  valley degeneracy, graphene has four Dirac cones at low energies (this is generally an {\em even} number for every purely 2D TR invariant system). Topological insulators are unique in this aspect, since an {\em odd} number of Dirac cones takes place to its surface states (while preserving TR symmetry). The most interesting consequence of this specialty is the strong linear momentum-spin coupling observed in the charge carries motion, what makes these dynamical states topologically robust against defects and impurities. Actually, the helical spin texture described by Hamiltonian (\ref{Dirac Hamiltonian}) present both charge current, $\vec j (\vec x)$, and spin densities, $\vec S(\vec x)$, lying only on the TI surface\cite{Lee 2009}, $ \vec j= v_F\Psi^\dag\vec\sigma\Psi\times \hat z = v_F \vec S \times \hat z$.

\begin{figure}[h]
\begin{center}
%\scalebox{.8} % Change this value to rescale the drawing.
\includegraphics[width=6.5cm]{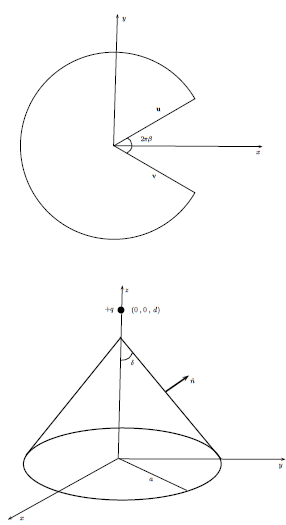}
\caption{ \label{conical geometry} The conical geometry: cut and removal of a sector $2\pi\beta$, $0 < \beta < 1$ from 2D plane; identification of lines $\mathbf u$ and $\mathbf v$ gives a cone; $\delta = \arcsin (1-\beta)$ is called the apperture angle, $\hat n$ is a normal to the surface). Note also the electric charge $q$ placed just above the apex which complete our system to be studied here.}
\end{center}
\end{figure}

In this paper, we show that TMEE and geometrical features interplay interetingly on a conical support rendering the TI the possibility of being electrically polarized only by controlling its geometry (apperture angle). For that we shall consider a 3D TI cone whose border is precisely its conical 2D surface, Fig.\ref{conical geometry}. We have analyzed the Hall current induced on the surface by an electric charge placed above the apex and we have also found a gravitational-like mechanism to reverse the cone polarization, say, narrower cones, with $\delta<30^o$ tend to accumulate positive charges around the apex region, while those with $\delta>30^o$ a net negative charge takes place there.

We start off by describing the conical geometry and the eletronic structure of such a topological insulator. 
Detaching a fraction of an Euclidian space generate a deficit angle $2\pi\beta$ and after identifying the edges 
a cone results (see Fig. \ref{conical geometry}). The cone surface makes an apperture angle, 
$\delta=\arcsin(1-\beta)$, with its symmetric $z$-axis. Although the cone is locally flat, once its curvature vanishes everywhere, global effects show up, for instance, when one goes around the tip (where curvature blows up; further details below). Conical geometry naturally emerges in (2+1)-dimensional Einstein gravity where a pointlike mass induces an angle (or area) deficit in the 2D spatial plane, playing a role analogous to the conical tip, accounting for the global lacking of space. Explicitly, the line element in this case reads
\be\label{metric}
ds^{2}= dt^{2}-dr^{2}-r^{2}\eta^{2}d\theta^{2},
\ee
with $0\leq \theta < 2\pi $ and $\eta=1-4GM$ ($G$ is the Newton's constant, which has dimensions of $length$, 
rather than $length^{2}$, in natural units, $h=c=1$). Note that although the situation looks trivial the coordinate $\theta$ ranges from $0$ to $2\pi \eta<2\pi$ indicating a angle deficit in this geometry. Then, the space 
part of the metric is that of a plane with a wedge removed and edges identified; the unique $2D$ spatial
geometry satisfying this description is the cone \cite{Staruszkiewikz63,Brown88}. Although no gravitons appear in this space-time, since Einstein tensor vanishes in the absence of energy-momentum, implying that gravity is experienced only locally, global effects coming from a massive particle are captured by the angle deficit and manifested by the conical tip
\cite{Staruszkiewikz63,Deser84,Giddings84,Vilenkin81}. 

Therefore, the dynamics of charge carriers in an ideal conical TI is equivalent to that of a (massless) Dirac 
particle in a gravitational field of a static particle of mass M in a $(2+1)$D space-time. This analogy suggests that Einstein gravitation in $(2+1)$ dimensions can be realized and experimentally probed in feasible condensed matter materials, such as in conical TI's systems (similar tests have been porposed with conical graphene sheets as w\cite{Lamert-Crespi,Furtado, Jakson}). 

\begin{figure}[h]
\begin{center}
%\scalebox{.8} % Change this value to rescale the drawing.
\includegraphics[width=6.5cm]{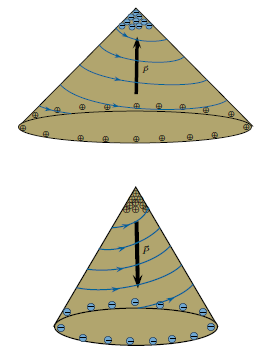}
\caption{\label{cone-polarization} Electric charge polarization in conical topological insulators: Wider TI cones (upper), with $\delta>30^o$ tend to accumulate negative charges (electrons) near the apex, while in narrower ones a net positive charge is expected to occurs there. Black arrows, $\vec{P}$, accounts for the cone polarization while the blue curves represent the Hall current flowing along the surface.}
\end{center}
\end{figure}

Let us start off by investigating how closed-path parallel transport around the cone tip affects Dirac spinors, $|\Psi\rangle$. Basically, it results 
in a kind of Aharonov-Bohm effect, but occuring without an electromagnetic field. Once it origins from the geometrical properties of the physical support (the cone) it is often referred to be {\em gravitational Aharonov-Bohm effect}. For clarify that, one solves the differential equations for the charge carries\cite{Burges85} (equivalently, we could make use of path-ordered products of the relevant affine connection\cite{Bezerra87}). After taking a complete round along the tip, wave-functions are changed according to:
\be
\label{ABphase}
|\Psi (2\pi)\rangle=\left(
  \begin{array}{cc}
    \exp(-i\phi) & 0 \\
    0 & \exp(i\phi)\\
  \end{array}
\right) |\Psi (0)\rangle,
\ee
with $\phi=\pi \sin\delta$. For instance, if $|\Psi (0)\rangle = p|+ \rangle + q|- \rangle$ ($p$ and $q$ are
constants), then, $|\Psi (2\pi)\rangle = p \exp(-i\phi)|+ \rangle + q \exp(i\phi)|- \rangle$ and, therefore, 
after the complete contour, the components up and down spins are mutually out of phase by 
$2\pi \sin\delta = 2\pi(1-\beta)$, which is exactly the angle remaining in the space. Hence, the angular deficit gives rise to a mismatch in the electrons and holes properties displacing on the conical TI surface, say, {\em up} and {\em down} components get a relative phase difference after transported around the tip (note that if $\delta=\pi/2$ we recover the usual plane, and no change appears, at all). This effect leads to
fluctuations of charge carriers density along the conical surface and may change several physical properties of the material. This is clearer realized and even detected in the induced superficial Hall current: When a vector like $\vec j$ is parallel transported around the conical apex its componentes change by\cite{Burges85}:

\begin{equation}
\left(\begin{array}{c}
         j_r(2\pi) \\
         j_\theta(2\pi)
       \end{array}\right)=
       \left(
         \begin{array}{cc}
           \cos(2\pi\gamma) &  \frac{1}{r_0\alpha}\sin(2\gamma)\\
            -r\alpha\sin(2\pi\gamma) & \frac{r}{r_0} \cos(2\pi\gamma)\\
         \end{array}
       \right)
       \left(\begin{array}{c}
         j_r(0) \\
         j_\theta(0)
       \end{array}\right)\,,
\end{equation}
with the initial condition $\theta=0$ at $r=r_0$ while and $\gamma=\sin\delta=1-\beta$ (see Fig. \ref{conical geometry}). Although Hall current induced by a charge, placed just above the cone apex, initially has only $\hat \theta$- component (for details see eq.(\ref{induced current}) below), whenever it circulates around the apex an extra term shows up along $\hat r$. Thus, the whole current is not closed once $j_r$ displaces towards either apex or cone botton (see below). Such an effect comes about in the conical geometry and identically vanishes in the usual flat plane. On the other hand, it may be faced as a (2+1)D gravity manifestation of a point mass and might be used to probe this theory at a microscopic scale.
As a explicit example let us consider the Hall current induced at the point $(r=r_0\,,\,\theta=0)$:
\be
\vec j (r_0\,,\,0)=j_\theta (r_0\,,\,0) \hat\theta\,,
\ee
which after complete a round on the apex will read:
\be
\vec j(r\,,\,2\pi)= \frac{\sin(2\pi\gamma)}{r_0\alpha} j_\theta (r_0\,,\,0)\hat r +
\frac{r \cos(2\pi\gamma)}{r_0} j_\theta (r_0\,,\,0)\hat \theta\,.
\ee
Observe that when $\gamma=1/2$, ($\delta=30^0$) the radial component of the current vanishes (this corresponds to $\beta=1/2$, say, a deficit equal to half of the 2D space). Interesting polarization-type effects appear for other $\gamma$ values: Indeed, if $\gamma<1/2$ ($\delta<30^0$, narrower cones) $J_r$ is positive, so that total current, $\vec{j}$, spirals towards cone botton. After a while cone will get a negative charge acumulated in the base (and a positive charge near the apex) what corresponds to an electric polarization along the $-\hat z$ axis (see Fig. \ref{cone-polarization}). On the other hand, for $\gamma>1/2$, ($\delta>30^0$, wider cones) the $j_r$ is negative, yielding a polarization directed from the botton to the apex. [Of course, the usual polarization brought about by charge $q$, placed just above the cone tip, remains independently. Actually, the net conical TI polarization may be also controlled by choosing the signal and magnitude of this charge and the magnetization direction, as accounted by eq. (\ref{induced current})].\\

Let us now study fields configuration brought about by that pointlike charge. Qualitatively, the in-plane component of the electric field generated by the point charge does induce a superficial and perpendicular current. This appears as a quantized Hall current as a physical manisfestation of topological magneto-electric 
effec (TMEE) and is given by:
\be \label{induced current}
\vec j=\pm \frac{e^2}{2h}\hat n\times\vec E=\pm\frac{q\, e^2}{4\pi\epsilon_0 h}\frac{[\,r\sin\alpha \,+\,(d-z)\cos\alpha\,]}{[\,r^ 2+(z-d)^2\,]^{3/2}}\hat\theta\,,
\ee
where $\pm$ accounts for the magnetization direction, $+$ for magnetization outward and $-$ for inward the surface, $\vec{M}=\pm\vec{M}_0 \hat{n}$; parameters $r$ and $z$ are restrited to the cone surface satisfying $z= c-\frac{r}{\tan \delta}$. The electric and magnetic fields induced by the electric point charge on the TI surface are determined through Maxwell equations augmented by TMEE\cite{Qi}, which comes about from the action below:
\ba
& S= &\int d^4x -\frac14 F_{\mu\nu}F^{\mu\nu} + \theta \frac{\alpha}{2\pi} \tilde{F}_{\mu\nu}F^{\mu\nu} \nonumber\\ & = &\int d^4x \Big(\frac12(|\vec{E}|^2 - |\vec{B}|^2) +\theta \frac{\alpha}{2\pi} (\vec{E}\cdot\vec{B})\Big)\,, 
\ea
which yields imediately:
\ba
& & \nabla\cdot \vec{D}=\nabla\cdot\vec{B}=0\,, \\
& & \nabla\times\vec{H} -\partial_t\vec{E}=\nabla\times\vec{E} -\partial_t\vec{B}=\vec{0}\,,
\ea
with the constitutive relations:
\ba
& &\vec D= \vec E +\vec P-2\alpha P_3(x)\vec B\,,\\
& &\vec H=\vec B-\vec M+2\alpha P_3(x)c \vec E\,.
\ea
$P_3(x)=\theta(x)/2\pi$ accounts for the magnetoelectric polarization, which vanishes in vacuum or in conventional insulators ($\theta=0$) and equals $\pm 1/2$ ($\theta=\pi$) in TI (the sign is determined by the direction of the surface magnetization \cite{Qi}). We have set $c=\epsilon_0=\mu_0=1$ and $\alpha$ is the fine structure constant.

The electric (magnetic) field outside the conical TI can be viewed as induced by
a point image electric charge (magnetic monopole) plus a line of image electric (magnetic) charge density
along the $z$ axis inside the cone. The image method can be used to find the quantitative expression
for the charge density. An altervative technique so-called {\em fractional image method} \cite{imagem cone} is more usefull
because the conical space can be viewed as a fraction of the Euclidian space. The results for the point 
charge electric above the apex cone in $z-$axes and in a arbitrary $(r\,,\,\theta\,,\,z)$ position outside the
cone are in progress and will be presented soon \cite{jakson progress}.

In summary, we have considered the magnetoelectric effect realized for  conical topological insulators. The Hall current, induced by a point charge, at the conical surface is obtained and present the interesting feature of providing conical electric polarization, say, the accumulation of negative (positive) charges, ${\cal Q}$, around the apex according the cone apperture angle: if $\delta>30^o$ we observe ${\cal Q}<0$ while for narrower cones, $\delta<30^o$ a positive ${\cal Q}$ takes place. This effect seems to be feasible of experimental verification, providing a microscopic manner of probind (2+1) dimensional Einstein gravity. The explicit solutions for the induced fields, along with possible geometrical effects tracked by them, are in order and will be communicated elsewhere\cite{jakson progress}.

The authors thank CNPq, FAPEMIG and CAPES (Brazilian agencies) for
financial support.


\begin{thebibliography}{2dspins}

\bibitem{Hsieh 2008} Hsieh, D., D. Qian, L. Wray, Y. Xia, Y. S. Hor, R. J. Cava, and M. Z.
Hasan, Nature \textbf{452}, 970 (2008).

\bibitem{Chen} Chen, Y. L., et al., Science \textbf{325}, 178 (2009).

\bibitem{Xia} Xia, Y., et al., Nature Phys. \textbf{5}, 398 (2009).

\bibitem{Chadov} Chadov, S., X. L. Qi, J. K\'ubler, G. H. Fecher, C. Felser, and S. C.
Zhang, Nature Mater. \textbf{9}, 541 (2010).

\bibitem{Wray} Wray, L. A., S.-Y. Xu, Y. Xia, Y. S. Hor, D. Qian, A. V. Fedorov, H.
Lin, A. Bansil, R. J. Cava, and M. Z. Hasan, Nature Phys. \textbf{6},
855 (2010).

\bibitem{TI review} For reviews, see X.L. Qi and S.C. Zhang, Phys. Today \textbf{63}, 33 (2010);
M.Z. Hasan and C.L. Kane, Rev. Mod. Phys. \textbf{82}, 3045 (2010) and X.L. Qi, S.C. Zhang,
Rev. Mod. Phys. \textbf{83}, 1057 (2011)

\bibitem{y. l. chen} Y. L. Chen, et al, Science \textbf{329}, 659 (2010).

\bibitem{nomura and nagaosa 2010} K. Nomura and N. Nagaosa, Phys. Rev. B, \textbf{82}, 161401(R) (2010).

\bibitem{Garate} I. Garate and M. Franz, Phys. Rev. Lett.
\textbf{104}, 146802 (2010).

\bibitem{vozmediano} M.I. Katsnelson, F. Guinea and M.H. Vozmediano, arXiv:1105.6132v2.

\bibitem{Liu} Q., Liu, C.-X. Liu, C. Xu, X.-L. Qi, and S.-C. Zhang, Phys.
Rev. Lett. \textbf{102}, 156603 (2009).

\bibitem{Lee 2009} D.H. Lee, Phys. Rev. Lett. \textbf{103}, 196804 (2009).

\bibitem{garate and franz 2010} I. Garate and M. Franz, Phys. Rev. Lett.
\textbf{104}, 146802 (2010).

\bibitem{Qi} X.-L. Qi, T. Hughes, and S.-C. Zhang, 2008, Phys. Rev. B \textbf{78},
195424 (2008).

\bibitem{Essim} A.M. Essin, J. E. Moore, and D. Vanderbilt, Phys. Rev. Lett.
\textbf{102}, 146805 (2009).

\bibitem{Fu} L. Fu, and C. L. Kane, 2008, Phys. Rev. Lett. 100, 096407.

\bibitem{LiuHamiltonians} C.-X. Liu, X.-L. Qi, H. Zhang, X. Dai, Z. Fang, and S.-C. Zhang, 
Phys. Rev. B \textbf{82}, 045122 (2010).

\bibitem{Castro-Neto} A. H. Castro Neto, F. Guinea, N. M. R. Peres, K. S. Novoselov, and
A. K. Geim, Rev. Mod. Phys. \textbf{81}, 109 (2009).

\bibitem{imagem cone} N. Engheta. ``Electrostatic ``Fractional" Image Methods for Perfectly 
Conducting Wedges and Cones" Departmental Papers (ESE) (1996). Available at: 
http://works.bepress.com/nader\_engheta/10

\bibitem{jakson progress} J.M. Fonseca, W.A. Moura Melo and A.R. Pereira, work in progress.


\bibitem{Brown88} J.D. Brown, Lower Dimensional Gravity, World
Scientific, New Jersey (1988).

\bibitem{Staruszkiewikz63}, A. Staruszkiewikz, Acta. Phys. Polon.
\textbf{24}, 734 (1963).

\bibitem{Deser84} S. Deser, R. Jackiw, and G. 't Hooft, Ann. Phys.
(N.Y.) \textbf{152}, 220 (1984).

\bibitem{Giddings84} S. Giddings, J. Abbott, and K. Kuchar, Gen.
Relativ. Gravit. \textbf{16}, 751 (1984).

\bibitem{Vilenkin81} A. Vilenkin, Phys. Rev. D \textbf{23}, 852
(1981).

\bibitem{Burges85} C.J.C. Burges, Phys. Rev. D \textbf{32}, 504
(1984).

\bibitem{Bezerra87} V.B. Bezerra, Phys. Rev. D \textbf{35}, 2031
(1987).

\bibitem{Lamert-Crespi}P.E. Lammert and V.H. Crespi, Phys. Rev. B {\bf 69}, 035406 (2004). 

\bibitem{Furtado} C. Furtado, F. Moraes and A.M.M. Carvalho, Phys. Lett. 
A \textbf{372}, 5368 (2008).

\bibitem{Jakson} J.M. Fonseca, W.A. Moura Melo and A.R. Pereira, Phys. Lett. 
A \textbf{374}, 4359 (2010).


\end{thebibliography}
\end{document}